# Graphene-protected copper and silver plasmonics

V. G. Kravets[1], R. Jalil[1], Y.-J. Kim[1,2], D. Ansell[1], D. E. Aznakayeva[1], B. Thackray[1], L. Britnell[1], B. D. Belle[1], F. Withers[1], I. P. Radko[3], Z. Han[3], S. I. Bozhevolnyi[3], K. S. Novoselov[1], A. K. Geim[1], and A. N. Grigorenko[1]

[1]*School of Physics and Astronomy, University of Manchester, Manchester, M13 9PL, UK*

[2]*Department of Chemistry, College of Natural Sciences, Seoul National University, Seoul, 151-747, Korea*

[3]*Institute of Technology and Innovation (ITI), University of Southern Denmark, Niels Bohrs Allé 1, DK-5230 Odense M, Denmark*

**Plasmonics[1,2] has established itself as a branch of physics which promises to revolutionize data processing[3,4], improve photovoltaics[5], increase sensitivity of bio-detection[6,7]. A widespread use of plasmonic devices is notably hindered (in addition to high losses) by the absence of stable and inexpensive metal films suitable for plasmonic applications. This may seem surprising given the number of metal compounds to choose from. Unfortunately, most of them either exhibit a strong damping of surface plasmons or easily oxidize and corrode[2]. To this end, there has been continuous search[8-10] for alternative plasmonic materials that are, unlike gold, the current metal of choice in plasmonics, compatible with complementary metal oxide semiconductor technology. Here we show that copper and silver protected by graphene are viable candidates. Copper films covered with one to a few graphene layers show excellent plasmonics characteristics surpassing those of gold films. They can be used to fabricate plasmonic devices and survive for at least a year, even in wet and corroding conditions. As a proof of concept, we use the graphene-protected copper to demonstrate dielectric loaded plasmonic waveguides[11] and test sensitivity of surface plasmon resonances. Our results are likely to initiate a wide use of graphene-protected plasmonics.**

Main obstacles to practical applications of plasmonic devices are the absence of chemically stable, relatively cheap and easily manufacturable plasmonic materials, nanofabrication issues and relatively high plasmonic losses. Metals are usually chemically active and their surfaces easily oxidize degrading their plasmonic characteristics. The surface plasmon resonance technique[6] – the most successful plasmonic application so far – relies on noble metals and, in particular, gold. Ubiquitous in modern plasmonics gold is however not compatible with standard silicon manufacturing processes, i.e., with complementary metal oxide semiconductor (CMOS) technology, due to very efficient diffusion of gold into silicon[12]. Various attempts have been made to search for alternative plasmonic materials, such as metal alloys, semiconductors and even superconductors[8,10,12]. Recently, graphene emerged as a viable candidate for plasmonic applications in the near infrared region of the spectrum[9,13-15]. However, a reliable and accessible plasmonic material for a wide range of practical applications remains wanted.

The problem of stability of a plasmonic material can be solved not only by searching for new candidates but also by finding a way to protect existing ones. For example, copper and silver exhibit higher conductivity than gold and could be excellent plasmonic materials[16,17,18] if only a way to protect them against oxidation is found[19]. To this end, a thin and inert coating impermeable to oxygen, water and other corroding agents is required. The coating should be sufficiently thin to allow one to use enhanced plasmonic near-fields. Graphene may provide such a coating[20,21]: it is one atom thick, mechanically strong, chemically inert and impenetrable to all gases and liquids[22]. Moreover, industrial-scale graphene sheets can now be grown by using chemical vapour deposition (CVD) and then transferred onto various surfaces where they show strong and uniform adhesion[20,23,24]. There have already been reports in which graphene was successfully employed to protect metals against corrosions and chemical reactions[21,25,26]. On the other hand, it has also been shown that copper with a graphene layer grown on top slowly deteriorates due to defects and cracks in graphene and within several months the copper surface becomes even worse than without protection[26].

In this report, we investigate the viability of graphene protection for plasmonic applications. We will concentrate on plasmonic geometries which utilize running plasmons and require a flat surface of metal with possible grooves and dielectric and/or metal loaded structures with an ultimate goal to develop active plasmonic elements where graphene would provide an electric control of optical signals. In particular, we show that copper films coated with transferred graphene exhibit the surface plasmon resonance (SPR) with phase sensitivity to chemical binding events of several orders of magnitude better than that of standard gold films. Graphene-protected copper does not display deterioration of the SPR over the studied period of one year, despite subjecting our samples to water and other chemicals. We have also fabricated a dielectric loaded surface plasmon polariton waveguide (DLSPPW)[11] and demonstrate its successful functioning. The graphene coating provides not only a corrosion barrier but also allows targeted (bio)functionalization of its surface. We illustrate the arising opportunities by using reversible graphene hydrogenation[7] and measuring the corresponding SPR sensitivity.

Copper thin films were made by electron-beam evaporation (Methods). Their morphology and optical properties varied relatively little with the film thickness and employed substrate [see Fig. 1(a)]. All fresh-deposited samples showed excellent plasmonic characteristics that gradually deteriorated with time, see below. To protect metal films against the deterioration, graphene was transferred on top by using the wet-transfer procedure[27]. We used both mechanically exfoliated graphene and graphene grown by CVD on copper (see Methods). A single layer of graphene transferred onto copper films [Fig. 1(b)] already showed excellent anticorrosion properties. However, to avoid corrosion through occasional defects and cracks, the second graphene layer was deposited, so that overlaps between defects present in different layers became highly unlikely. We could completely seal the metal surface with using more layers (3-5) but in practice two and often even one protective layers were found sufficient to preserve plasmonic properties for the whole duration of our studies (~ 1 year), in contrast to graphene grown directly on copper[26].

To assess plasmonic properties of graphene-protected Cu, we have studied the SPR by using the Turbadar[28]-Kretschmann-Raether ATR geometry schematically shown in Fig. 1(c) and propagation of plasmon-polaritons in the DLSPPW geometry shown in Fig. 1(d). The SPR for two Cu films (as fabricated; unprotected) are shown in Fig. 2(a) and (b). For the Cu film of thickness $d$ =43.5 nm the strongest resonance is observed at the angle $\theta_R$ = 49.5° for the wavelength $\lambda_R$ = 588 nm; the prism refractive index $n$ was 1.5. The minimum reflection for the SPR curve is $\Psi_{min}$ = 0.7° (for the ellipsometric parameter) which translates into the minimum intensity reflection $R_{min}$ = 5×10$^{-4}$ (Fig. 2(b)). The quality factors are $Q$ = 19 and 12, if extracted from the ellipsometric and reflection curves, respectively. Here $Q = \lambda_R / (\Delta\lambda_{FWHM})$ where $\lambda_R$ is the resonance wavelength corresponding to the SPR minimum and $\Delta\lambda_{FWHM}$ is the full-width of the resonance at half-minimum of the resonance curve. These characteristics are better than those observed for Au films with the SPR at the same wavelength[10,29], which can be attributed to the higher conductivity of copper and better morphology of our evaporated films.

If left exposed to air or even placed in a desiccator or low vacuum, the unprotected Cu samples have oxidized rapidly and the quality of SPR deteriorated; see Figs 2(c) and (d). After 30 days, the deepest SPR curve shifted to $\theta_R = 53°$ and the resonance minimum was observed at $\lambda_R = 637$ nm. The minimum reflection increased by a factor of 20 to $\Psi_{min} = 11.5°$ and the minimum intensity by a hundred times to $R_{min} = 4\times10^{-2}$, whereas the quality factor of the SPR curve dropped to $Q = 7$ and 6 for the ellipsometric and reflection curves, respectively. In contrast, the samples protected by graphene did not show any degradation of plasmonic characteristics and in the SPR curve even after one year (Fig. 2). Moreover, graphene coating even improves the SPR quality, if graphene is transferred within minutes after exposing Cu films to air (which is most probably connected to the decrease of oxidation time). In this case, Figs 2(e) and (f) show that for the graphene-covered part of the same Cu film with $d = 43.5$ nm, the deepest SPR occurs at $\theta_R = 49°$ and $\lambda_R = 603$ nm and exhibits $\Psi_{min} = 0.4°$ and $R_{min} = 3\times10^{-4}$. $Q$ also increases to 20 and 13 for the ellipsometric and reflection curves, respectively. We stress that such small values of $R_{min}$ could not previously be achieved for Au films[6], because of their morphology that has proven difficult to improve and control[30]. At the same time, the reflectance $R_{min}$ is important for phase sensitive biosensing[7], and special efforts have been made to improve it by using collective plasmon resonances[31] and topological darkness[7]. For completeness, Figure 3(a) provides a comparison of SPR curves measured in air for a fresh copper film, a film protected by one graphene layer and a film protected by 2 graphene layers. Notice that the each graphene layer induces a small red shift of ~ 10 nm in the SPR position.

Main applications of the SPR lie in biosensing which necessitates a wet environment. Therefore, it is important to check that graphene is able to protect Cu films in water. Figure 3(b) shows a SPR curve for an unprotected film measured in water immediately after evaporation. The SPR is observed at at $\theta_R = 60°$ and $\lambda_R = 664$ nm ($n = 1.8$ in this case). The unprotected copper corrodes quicker in water than air, and the SPR notably degrades in just 24 hours; see

Fig. 3(c). One can see that $\Psi_{min}$ increases from 1° to 21° and $Q$ drops from 19 to 9.5. On the contrary, graphene-protected films do not show deterioration in plasmonic properties in water over the whole period of our studies. Figure 3(d) shows the typical SPR curves measured in water for a Cu film protected by a single layer of graphene. In water, graphene yields a red-shift of 115 nm (which is connected with the usage of SPR resonance of longer wavelengths that becomes more pronounced after graphene transfer) with a simultaneous decrease in $Q$. At the same time, the minimum reflection in the SPR curve drops to $\Psi_{min} = 0.3°$. These changes can be well described by the Fresnel theory taking into account partial oxidation of the Cu films (see the inset of Fig. 3(a), Fig. 4(c) and Methods).

The success of using graphene as a protective layer for Cu poses the obvious question whether graphene coating can be employed for other metals. For example, silver is well known as one of the best possible plasmonic materials[10] but it also notoriously known for its fast oxidation and degradation. There were earlier attempts to apply graphene coating to safeguard plasmonic response of silver[32] which found that the SPR response is stable in time but unfortunately it is also mediocre. We have identified the reason for the invasive influence of graphene which lies with a graphene transfer protocol and developed a new graphene transfer procedure which solved this problem (see Methods). As a result, we were able to protect Ag films with graphene without deteriorating the plasmonic properties of silver. Figures 3(e) and (f) show SPR curves for a fresh Ag film and an Ag film covered with a monolayer of CVD graphene, respectively. One can see that graphene does not deteriorate the excellent SPR response of silver: the resonance minimum changes from $\Psi_{min} = 0.3°$ for the freshly prepared Ag film, see Fig. 3(e), to $\Psi_{min} = 0.7°$ for graphene-protected Ag film, Fig. 3(f). We briefly compare the plasmon propagation length in Au, and graphene-protected Cu, Ag. Using optical constants extracted from ellipsometry, we estimate the following numbers for the propagation length assuming that the films are complete flat: Au – 11.9 μm, Cu – 8.2 μm and Ag – 21.2 μm at the wavelength of 630nm. The morphology of the films reduces these numbers to about 3-4 μm for our Au and graphene-protected Cu, while the propagation length is about 10 μm for Ag.

Having established that graphene coatings efficiently protect plasmonic properties of copper and silver films, we have measured sensitivity of our SPR devices to binding chemicals, which is one of the most important parameters for applications. To this end, we have chosen to use reversible graphene hydrogenation (Methods). Sensitivity to local environment (local refractive index) was also studied by utilizing glycerol-water mixtures. Figs 4(a) and (b) show evolution of ellipsometric parameters when samples were exposed to atomic hydrogen, which led to partial hydrogenation of the graphene layer. The hydrogenation ratio was monitored by Raman spectroscopy [inset of Fig. 4(b)] through the ratio of the intensity of the G and D peaks[7]. The hydrogenation level after 30 min of the exposure was ≈17% which corresponds to the areal mass density $\sigma$ of adsorbed hydrogen of ~10 pg/mm$^2$. Accordingly, the mass sensitivity of our graphene-protected SPR sensor was $\delta\lambda/\sigma$ ~0.15 nm and $\delta\Delta/\sigma$ ~30° per pg/cm$^2$ in terms of wavelength and phase measurements, respectively. Assuming a minimal detectable phase shift of $5\times10^{-3}$ degrees[33], we obtain the minimal detectable adsorbed mass of ~ 0.2 fg/mm$^2$. This detection limit is similar to that of plasmonic metamaterials with topological darkness[7] and is 4 orders of magnitude better than the sensitivity achieved for gold-based plasmonics[6].

Figures 4(c) and (d) illustrate the sensitivity of our SPR sensors for bio-applications where changes in local environment were mimics by replacing water (*n*=1.33) with a water-glycerol mixture (*n*=1.34). This charge results in a pronounced red-shift of the SPR, which translates into the wavelength sensitivity $\delta\lambda/\delta n \sim 5600$ nm/RIU, where RIU is the refractive index unit, and $\delta\Delta/\delta n > 2\times10^4$ deg/RIU for the phase measurements. The former sensitivity is comparable with the theoretically envisaged sensitivity for gold-based SPR sensors[34] (~$10^4$ nm/RIU), whereas the latter is significantly better due to smaller reflection in the SPR minimum. It is worth mentioning that graphene and graphene oxide show good biocompatibility[35,36].

Finally, we show that graphene-protected copper can also be used in complex plasmonic devices that require nanofabrications (for example, in waveguides and resonators). To this end, Figs 5(a) and (b) show schematics and an optical micrograph of a plasmon-polariton waveguide that we have chosen as a test structure. The waveguide consists of a DLSPPW[11] made on top of graphene-protected copper and the coupling and de-coupling gratings (see Methods). We excited the waveguide by illuminating the coupling grating with a He-Ne laser ($\lambda =632.8$ nm) and observed radiation coming out from the decoupling grating [Fig. 5(c)]. Figure 5(d) shows the dependence of the transmitted light intensity as a function of the waveguide length. This yields the decay length for propagating waveguide modes of 10 μm, which suggests a strong contribution from photonic modes. While the detailed mode analysis of the fabricated DLSPPW and its waveguiding characteristics along with the excitation efficiency is still to be carried out, the fact of radiation transfer between the in- and out-coupling gratings is well established [Fig. 5(c)]. It is important to mention that Cu films without graphene protection would not survive the nanofabrication procedures that take time and involve baking steps speeding up Cu corrosion.

To conclude, graphene-protected copper is a viable alternative to the conventional noble metals in plasmonics applications. It is relatively cheap, stable, reproducible and high-quality plasmonic material that is suitable for nanofabrication and CMOS compatible. The graphene-protected silver provides an exciting opportunity for applications which require low plasmonic losses. With the current rapid progress in CVD growth of hexagonal boron nitride and other 2D crystals, we envisage that they can also be used as a quality protective coating for plasmonics.

**Methods.**

**Film depositions.** The copper films were produced by electron-beam evaporation at a base pressure of about $10^{-7}$ mbar and growth rate of 0.3 nm/s (film thickness was monitored by calibrated quartz microbalance). As an electron-beam target, we used 99.99% Cu from Sigma-Aldrich. A thin adhesion layer of Cr with thickness of about of 1.5 nm was evaporated onto a substrate before copper. Glass substrates of sizes 25 mm x 25 mm and thickness of 1 mm were used for all the studied samples. The substrates were ultrasonically cleaned in heated acetone and isopropanol before deposition. In principle, CVD graphene can be grown directly on copper films. However, we found that plasmonic properties of copper with CVD grown graphene are quite poor due to large surface roughness produced by CVD process. For this reason, in order to produce graphene-protected copper we have used transfer procedure described below. This procedure (as well as CVD graphene) is inexpensive and can be easily automated.

**Growth of CVD graphene.** Cm-size graphene films were grown on Cu by using the CVD method[24]. A 25-µm thick Cu foil was placed inside a quartz tube and then heated to 1000°C with a $H_2$ flow at rate of 20 cm$^3$/min and a pressure of 200 mTorr. To remove the native oxide layer, the foil was first annealed at 1000°C for 30 minutes. Then a gas mixture of $H_2$ and $CH_4$, with flow rates of 20 and 40 cm$^3$/min, respectively, was introduced into the chamber. CVD growth was performed at a pressure of 600 mTorr for 30 mins. Finally, the CVD chamber was rapidly cooled to room temperature in hydrogen atmosphere. The grown films were predominantly single-layer graphene without many defects, verified by Raman spectroscopy.

**Graphene transfer procedure.** CVD-grown graphene was transferred on to the target sample by using the following procedure[7,27]. First, graphene-on-Cu was covered by poly(methyl methacrylate) (PMMA) by using spin coating. Then the PMMA film with graphene layer attached was isolated by chemically etching away the Cu foil. This PMMA-graphene film was then transferred onto the target copper films into a desired position under an optical microscope. Finally, the PMMA layer was removed in acetone and the graphene surface was further cleaned by hydrogen annealing at 200°C for 60 mins.

CVD graphene transfer on silver was performed as follows. Graphene grown on copper foil was removed by etching the copper in 0.1 Mol ammonium persulfate for ~6 hours. Before etching, graphene was covered by 400nm of PMMA with a tape window cut out and placed on the PMMA so that the membrane can be mechanically moved after etching. The free floating graphene membrane was transferred to a clean dish of deionised (DI) water for 5 min then to a second dish of DI water for a further 5 min to remove contamination from etchant solution. The graphene membrane was then taken out of the DI water and left to dry in ambient conditions for about 10 min. A droplet of IPA was placed on the Ag substrate and the CVD graphene was placed on the IPA. The evaporating IPA helps the graphene adhere to the Ag without the need for heating. (Ag can be easily oxidized in water and in air if the temperature is raised above ~50 degrees.) The sample was then placed in vacuum (~$10^{-5}$ mbar) for about 30 min to pump away any remaining liquid from the interface and then the sample was heated to 50 degrees for half an hour to soften the PMMA. The PMMA was then dissolved in acetone, the sample was rinsed with IPA and nitrogen dried.

**Graphene hydrogenation.** In order to bind hydrogen to the graphene surface we used a cold hydrogen dc plasma at a low-pressure (~0.1mbar) $H_2$/Ar (1:10) mixture. The plasma was ignited between Al electrodes. The level of hydrogenation was estimated by measuring the $D$ to $G$ peak intensity for hydrogenated samples registered by using a Renishaw RM1000 spectrometer at a 514nm excitation wavelength. A typical distance between hydrogen sites $L_D$ was calculated as $L_D = 4.24 \times 10^{-5} \lambda^2 \sqrt{I(G)/I(D)}$, where $\lambda$ is the wavelength measured in nanometers, $I(G)$ and $I(D)$ are the intensities for the $G$ and $D$ Raman peaks of hydrogenated graphene[7]. This yields $L_D \approx 10$ nm after the 30 min hydrogenation. We assume that the size of possible hydrogen clusters is smaller than the inter-cluster distance (~5nm), which gives us an estimate of 17% hydrogenation after 30 min. It is worth noting that graphene hydrogenation is a reversible process which can be reverted by a soft anneal[7].

**Fabrication and characterization of DLSPPW.** Dielectric waveguide structures were produced by photolithography on graphene-covered Cu films by using the negative tone photoresist ma-N 1405 from Micro-resist Technology. The substrate was pre-heated to 200°C in order to improve adhesion of the resist. We used a 250-nm thick resist layer spin-coated on to the sample and patterned it with the Microtech Laserwriter LW405 system. The pattern was developed in ma-D 533/S developer and rinsed in deionised water. Arrays of waveguides of width 1.0 μm and lengths 10 μm, 20 μm and 30 μm were fabricated. The pitch of coupling and de-coupling gratings was 1.2 μm; the grating ridge width was 1.0 μm.

To study light propagation in the waveguides, we illuminated the waveguide coupler with a 632.8 nm He-Ne laser at normal incidence by using a 100x objective lens to give a spot size of ~500 nm. The incident laser power was ~0.5 mW. The transmitted light intensity was evaluated from calibrated CCD images.

**Theory.** The reflection of our devices has been calculated by using the Fresnel theory which takes into account light refraction on the right-angle prism and light reflection from a multi-layered structure facing air or glycol-water mixtures. The thickness of copper and Cr layers were taken from microbalance measurements and checked by spectroscopic ellipsometry. The optical constants for all the layers were also measured by spectroscopic ellipsometry. Graphene optical constants were taken in the analytical form and taking into account contributions from both intra and interband transitions. The only fitting parameter was the thickness of an additional layer of copper oxide (<0.5 nm) that is formed immediately after the sample is taken out of the deposition chamber.

**Acknowledgements**

We are grateful for the support by SAIT GRO Program and EPSRC grant EP/K011022/1. Y.-J. Kim was supported by the Global Research Laboratory Program (2011-0021972) of the Ministry of Education, Science and Technology, Korea.

**Authors Contributions**

ANG and AKG conceived the idea. VGK, RJ, Y-JK, BDB, DA made the devices. VGK, BT, LB, KSN, ANG, AVK modified samples and performed measurements. IPR, ZH, SIB designed waveguides. DEA performed calculations. All the authors contributed to discussion of the project. ANG guided the project.

**Additional information**

Correspondence and requests for materials should be addressed to ANG.

**Competing financial interests**

The authors declare no competing financial interests.

**Figure Captions.**

Fig. 1. **Testing graphene-protected Cu as a plasmonic material.** (a) Refractive index of copper films extracted by using spectroscopic ellipsometry. (b) Schematic of a typical graphene-protected sample. (c) ATR scheme for surface plasmon resonance measurements. (d) Schematics of a dielectric loaded plasmon polariton waveguide.

Fig . 2. **Comparison of the SPR in protested and non-protected copper films.** (a) SPR ellipsometric reflection $\Psi$ for a fresh unprotected sample as a function of wavelength ($\tan(\Psi)\exp(i\Delta) = r_p / r_s$, where $r_p$ and $r_s$ are reflection coefficients for *p*- and *s*-polarizations, respectively). The thickness of the copper film was 43.5 nm. (b) *p*-polarized intensity reflection $R_p$ for the same sample as a function of wavelength. (c, d) Same measurements as in (a) and (b), respectively, after 30 days. The inset in (c) emphasizes the degradation of the SPR with time. (e, d) Same as the above but for graphene-protected Cu and half a year.

Fig. 3. **Stability of graphene-protected SPR.** (a) Spectral dependence of the SPR measured at 49° in air for freshly fabricated unprotected Cu (black curve), Cu protected with a single graphene layer (red) and Cu with a double layer protection (blue). *d* = 43.5 nm. The inset shows results of our modelling for this experiment. (b) fresh unprotected sample in contact with water. (c) same as (b) after 24 hours in water. (d) graphene-protected sample in water after half a year. A large SPR shift compared to (b) is connected with the usage of SPR resonance of longer wavelengths which becomes

more pronounced after graphene transfer. (e) SPR ellipsometric reflection for a fresh unprotected silver film ($d$ = 60 nm). (f) The same film covered with graphene.

Fig. 4. **Sensitivity of copper SPR.** (a) SPR curves at different levels of graphene hydrogenation. (b) Spectral dependence of the ellipsometric phase $\Delta$ for different hydrogenation levels. The inset shows the corresponding Raman spectra. (c) SPR curves in two different liquids. The coupling prism has refractive index $n$=1.8. The inset shows results of our modelling for (c). (d) The spectral dependence of the phase for the two liquids.

Fig. 5. **Waveguides using graphene-protected copper.** (a) Schematics of DLSPPW. (b) Microscopy image of the fabricated waveguides. (c) CCD image of the light propagation along the waveguide. (d) The dependence of the transmitted light intensity on waveguide's length.

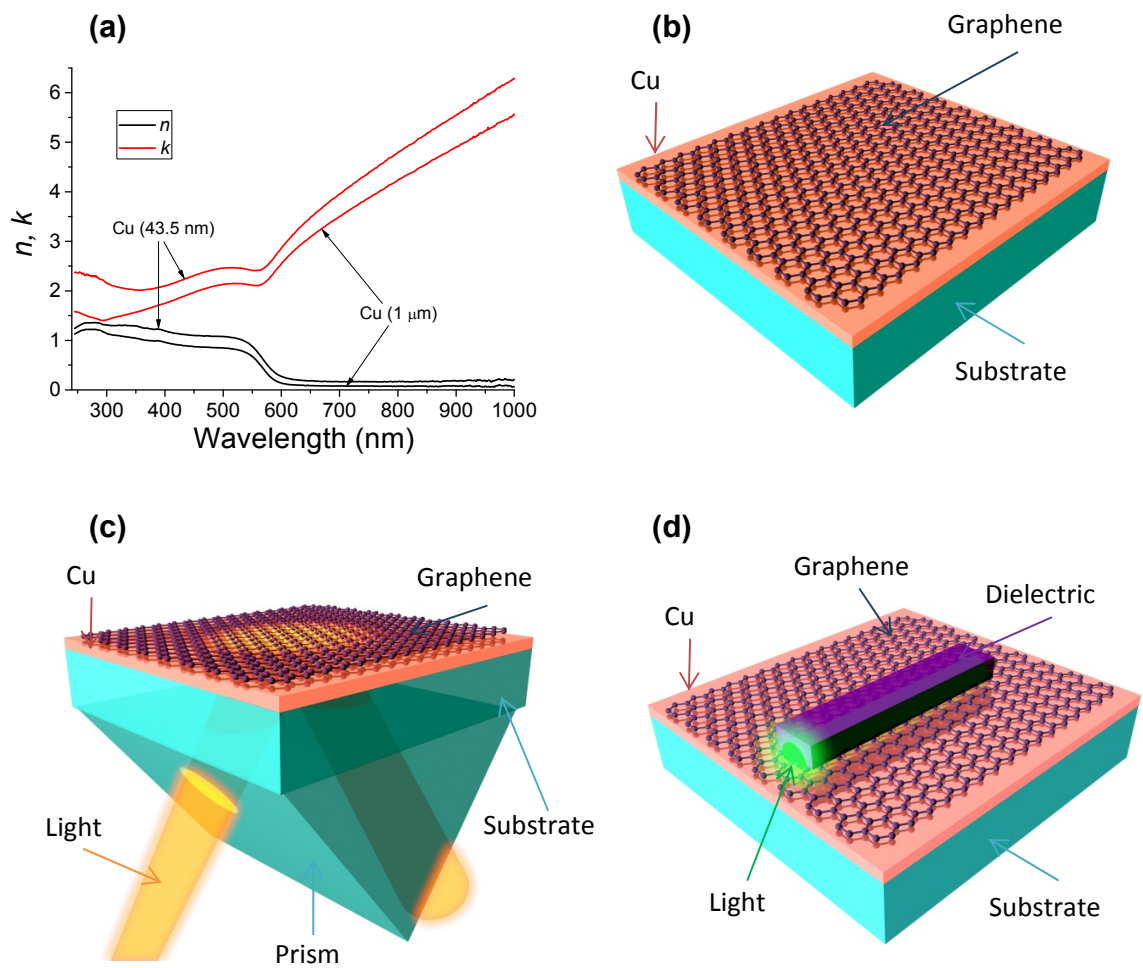

Figure 1

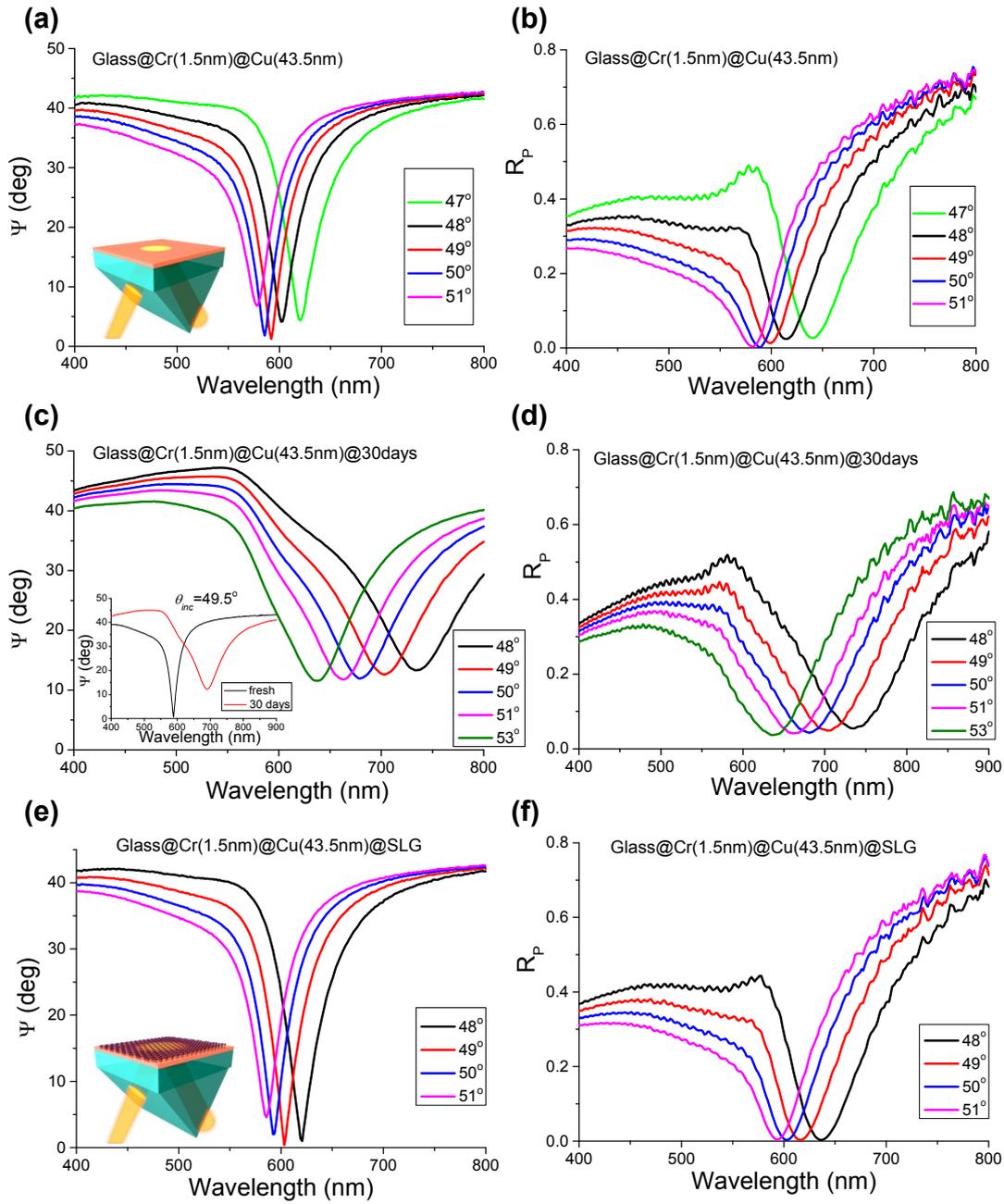

Figure 2

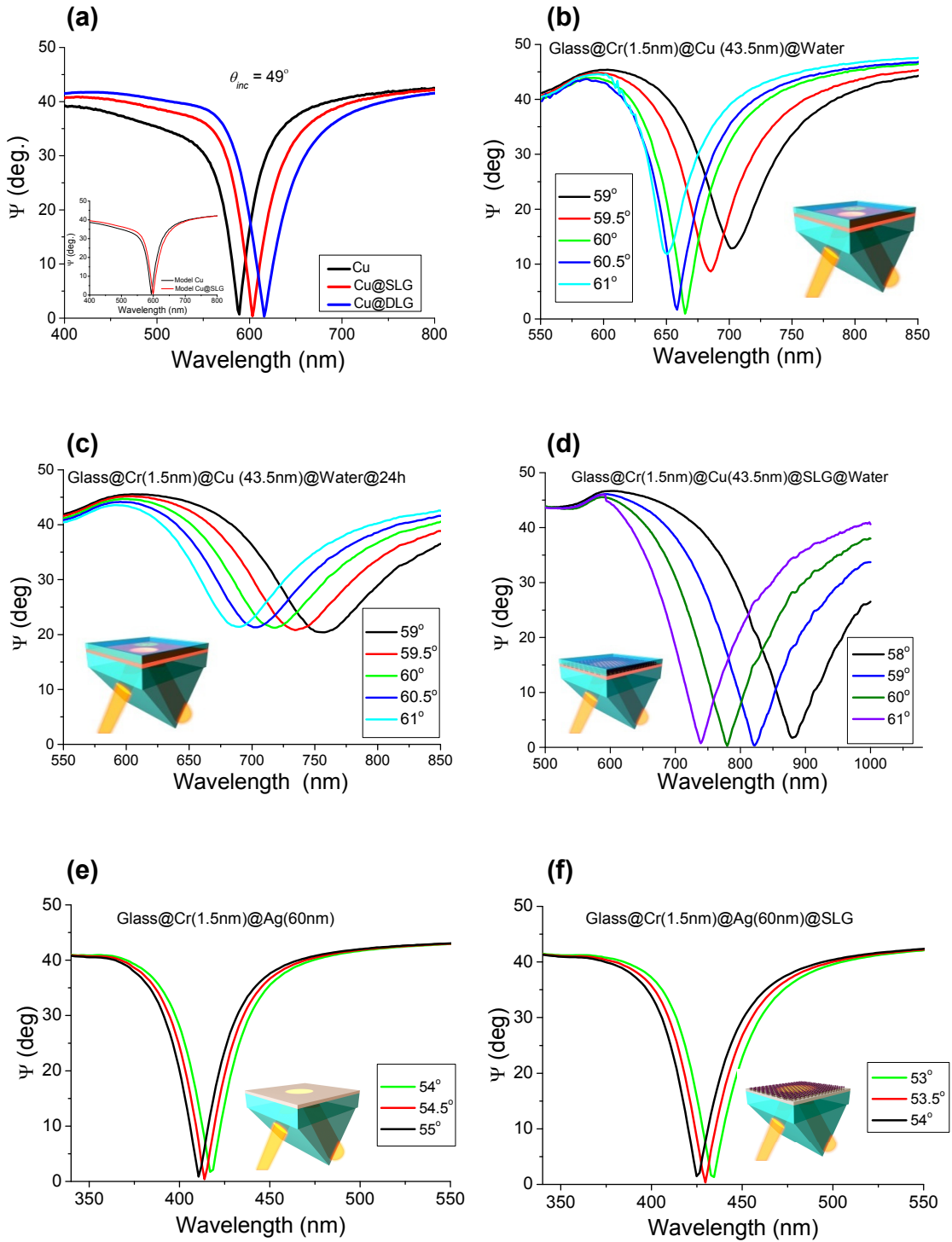

Figure 3

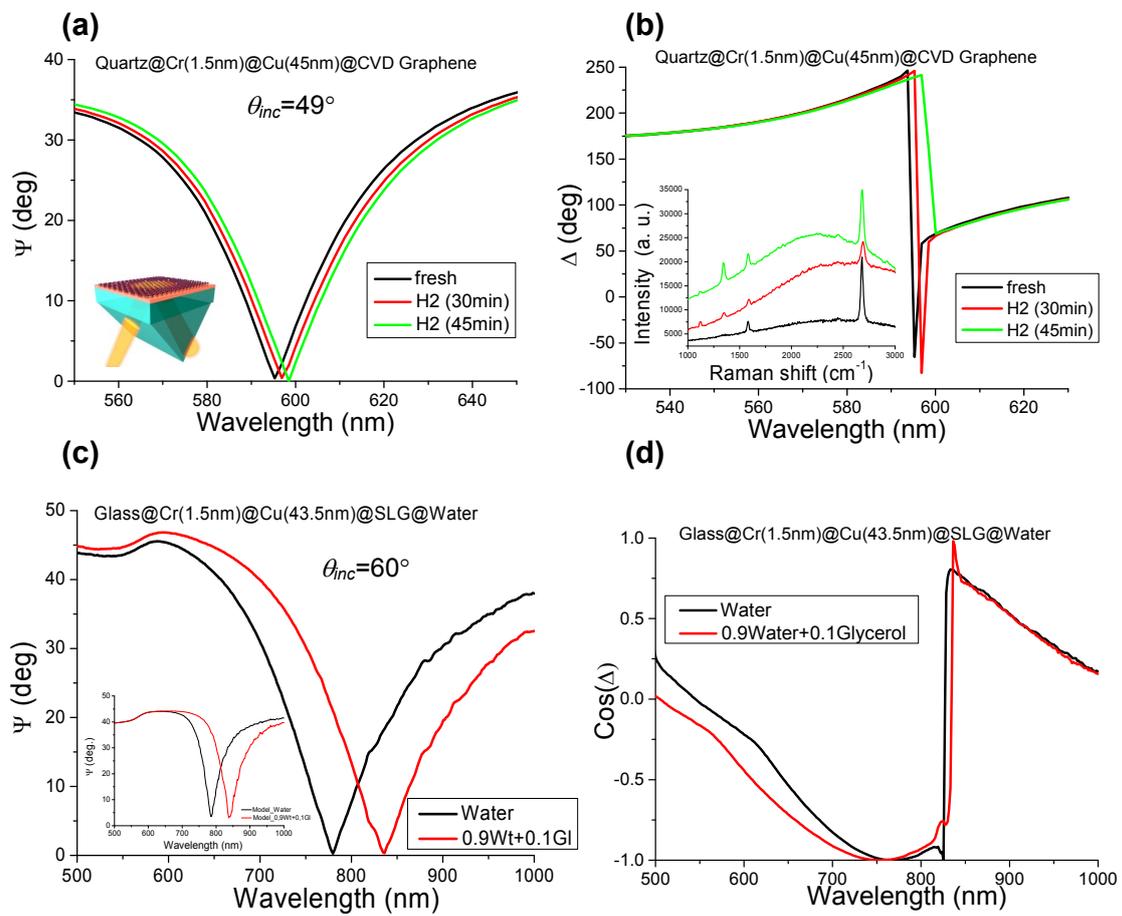

Figure 4

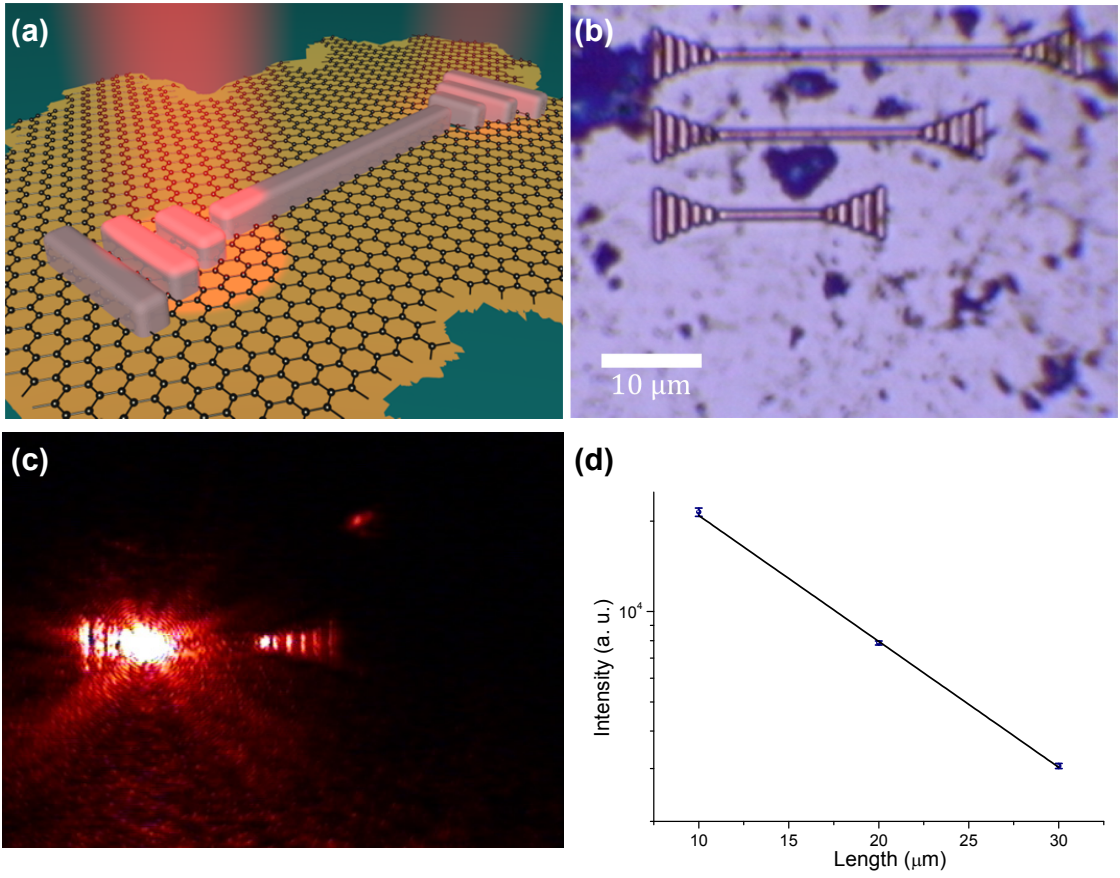

Figure 5